\documentstyle[12pt,cite]{article}   %%LaTeX 2.09

\newcommand{\non}{\nonumber}
\def\lequiv{\raise 0.4ex \hbox{$<$} \kern -0.8 em \lower 0.62 ex \hbox{$\sim$}}
\def\gequiv{\raise 0.4ex \hbox{$>$} \kern -0.7 em \lower 0.62 ex \hbox{$\sim$}}
\newcommand{\be}{\begin{equation}}
\newcommand{\ee}{\end{equation}}
\newcommand{\bea}{\begin{eqnarray}}
\newcommand{\eea}{\end{eqnarray}}
\newcommand{\smallz}{{\scriptscriptstyle Z}} %  a smaller Z
\newcommand{\smallw}{{\scriptscriptstyle W}} %
\newcommand{\smallh}{{\scriptscriptstyle H}} %
\newcommand{\mz}{M_\smallz}
\newcommand{\mw}{M_\smallw}
\newcommand{\mh}{M_\smallh}
\newcommand{\dr}{\mbox{$ \Delta r$}}
\newcommand{\equ}[1]{Eq.\,(\ref{#1})}

\def \mt   {M_t}
\def \gev  {\mbox{ GeV}}
\def \seff {s^2_{eff}}
\def \sineff {\sin^2\theta_{eff}^{lept}}
\def \ms   {\overline{\mbox{MS}}}

\def \ew   {electroweak\ }
\def\pl#1#2#3{{\it Phys. Lett. }{\bf B#1~}(19#2)~#3}
\def\zp#1#2#3{{\it Z. Phys. }{\bf C#1~}(19#2)~#3}

\def\pr#1#2#3{{\em Phys. Rev. }{\bf D#1~}(19#2)~#3}
\def\np#1#2#3{{\em Nucl. Phys. }{\bf B#1~}(19#2)~#3}

\begin{document}              
\begin{titlepage}
\begin{flushright}
        \small
        TUM-HEP-342/99\\
        MPI-PhT/99-06\\
        KA-TP-3-1999\\
%        hep-ph/9903249\\
        March 1999
\end{flushright}

\begin{center}
\vspace{1cm}
{\Large\bf Test of  the Heavy Top Expansion in the Evaluation of $\mw$ and
$\sineff$ 
%\footnote{version of 5.2.99 including the last changes offered ro Georg}
}

\vspace{0.5cm}
\renewcommand{\thefootnote}{\fnsymbol{footnote}}

{\bf           P.~Gambino$^a$,
                        A.~Sirlin$^b$\footnote{Permanent address: New York
                        University, Physics Dept., New York, NY  10003, USA}, and 
G. Weiglein$^c$}
\setcounter{footnote}{0}
\vspace{.8cm}

{\it
        $^a$ Technische Universit{\"a}t M{\"u}nchen,\\
                       Physik Dept., D-85748 Garching, Germany \\
\vspace{2mm}
        $^b$ Max Planck Institut f{\"u}r Physik (Werner-Heisenberg-Institut),\\
        F{\"o}hringer Ring 6, D-80805 M{\"u}nchen,  Germany \\
\vspace{2mm}
$^c$ Institut f\"ur Theoretische Physik, Universit\"at Karlsruhe,\\
D-76128, Karlsruhe, Germany}

\vspace{1.5cm}

{\large\bf Abstract}
\end{center}
In order to test the accuracy of the Heavy Top-mass Expansion (HTE) employed in
recent two-loop 
calculations of $\mw$ and $\sineff$, we consider their  contributions
to subtracted quantities of the form $(\mw)_{sub}=\mw(\mh)-\mw(\mh^0)$,
where $\mh^0$ is a reference point. The results are compared with those
obtained by a precise numerical evaluation  of 
all the  two-loop contributions  involving  both the Higgs boson and 
a fermion loop. 
For the choice $\mh^0=65$ GeV, and over the large range 
65~GeV\,$\le\mh\le1$ TeV, we find very small differences between the precise 
and HTE calculations, amounting to
$|\delta\mw|\le0.8$ MeV and $|\delta\seff|\le 1.2 \times 10^{-5}$.
Although corrections involving light fermions are necessary for the consistency
and test of existent calculations, we also discuss the separate contributions
from the top-bottom isodoublet. In this case, 
the differences are larger, although still small, namely 
$|\delta\mw|\le1.9$ MeV and $|\delta\seff|\le 4.5 \times 10^{-5}$.

\noindent

% PACS numbers: 

\end{titlepage}
\newpage

%%%%%%%%%%%%%%%%%%%%%%%%%%%%%%%%%%%%%%%%%%%%%%%%%%%%%%%%%%%%%%%%%%%%%%%%%%%%%%%
The corrections of $O(g^4 \mt^2/\mw^2)$, evaluated
on the basis of Heavy Top-mass Expansion (HTE) techniques,
are now incorporated in the 
 calculation of the \ew observables $\mw$, $\sineff$, and $\Gamma_f$ 
with $f\neq b$  \cite{dgv,dgs,dg}. In these papers, the HTE is applied 
in two different ranges of the Higgs boson mass $\mh$, and it is found that 
both expansions match nicely at $\mh\approx\mw$.
The results have reduced significantly the scheme and scale dependence of the
overall corrections, and have decreased the estimated Higgs-mass bounds by
$\approx $ 30\% \cite{dgps}.

The dependence of the electroweak observables $\mw$, $\sineff$, and $\Gamma_f$
on $\mh$ was also studied recently in Refs.\cite{bw,georg}. 
In these calculations, all
the two-loop contributions involving both a fermion
loop and the Higgs boson ($H$) have been taken into account accurately, i.e. 
without using the HTE, by a combination of algebraic and precise numerical
methods. For brevity we will refer to this class of diagrams as 
${\cal C}^{(2)}(f,H)$.
We note that graphs of this class in which the $H$ couples to external
fermions are at most of $O(M_\mu^2/\mw^2)$ and, therefore, negligible. As not
all of the two-loop contributions are included, the diagrams in 
${\cal C}^{(2)}(f,H)$
are actually divergent. However, the divergences are $\mh$-independent, so that
it is possible to evaluate accurately the contributions of 
${\cal C}^{(2)}(f,H)$ to subtracted corrections, such as 
$(\dr)_{sub} (\mh)\equiv \dr(\mh) - \dr(\mh^0)$.
Here $\dr$ is the correction introduced in Ref.\cite{si80} 
and $\mh^0$ is a reference value. ${\cal C}^{(2)}(f,H)$
 includes all the $\mh$-dependent contributions enhanced by
factors $(\mt^2/\mw^2)^2$ ($n=1,2$), as well as all remaining  
$\mh$-dependent effects involving the top quark or a light fermion loop. 
On the other hand, since 
two-loop purely bosonic self-energy contributions, as well as two-loop boxes
and vertex parts, are not included, we stress that ${\cal C}^{(2)}(f,H)$
does not contain the full two-loop $\mh$-dependence, a limitation that also 
applies to the calculations of Refs.\cite{dgv,dgs,dg,dgps}. 
We also point out that, in the approach of Refs.\cite{bw,georg}, the 
contributions of ${\cal C}^{(2)}(f,H)$ to subtracted radiative corrections
must be calculated at constant values of $\mw$, rather than $\mw(\mh)$, 
in order to  ensure the cancelation of divergences. The terms neglected 
in this approximation are formally of three-loop order and are expected to 
induce only a small error in the finite parts.

The aim of this paper is to test the precision of the HTE, as  employed in
Ref.\cite{dgv,dgs,dg}, and the resulting accuracy in the evaluation of 
${\cal C}^{(2)}(f,H)$. 
With this objective in mind, we consider the $O(g^4)$
contributions from Ref.\cite{dgs} to subtracted  corrections, relevant to the
calculation of $\mw$ and $\sineff$, and compare the results with those obtained
for the same quantities
 from the accurate two-loop calculation of ${\cal C}^{(2)}(f,H)$.
In order to facilitate the comparison, we employ OSII,
one of the on-shell schemes of Ref.\cite{dgs}, since  Ref.\cite{bw,georg} 
employs also the on-shell scheme of renormalization. 
We also avoid, as much as possible, deviations arising from different
treatments of higher-order corrections not contained in ${\cal C}^{(2)}(f,H)$.
QCD corrections are
excluded in both calculations, as they do not play a significant role in the
test of the HTE.

 We first note that Ref.\cite{dgs} employs the conventional framework
 \cite{si84} 
\bea
&& \ \ \ \ \ \ \ \ \ \ \ \ \
s^2 \, c^2 = \frac{A^2}{\mz^2} \frac1{(1-\Delta r)}\ ,\non\\
&&c^2\equiv 1-s^2 = \mw^2/\mz^2\, ,  \ \ \ A^2= \pi\alpha/\sqrt{2} G_\mu\ ,
\label{den}
\eea
while Ref. \cite{bw} uses the alternative expression
\be
s^2 \, c^2 = \frac{A^2}{\mz^2} (1+\Delta r_N).
\label{num}
\ee
The subscript $N$ reminds us that, in \equ{num}, $\dr$ has been 
introduced in the numerator.  Through two-loop order, \equ{den} leads to 
\be
\frac1{1-\dr}= 1+\dr^{(1)} + (\dr^{(1)})^2 + \Delta\tilde{r}^{(2)},
\label{drexp}
\ee
where $\dr^{(1)}$ is the original one-loop result of Ref.\cite{si80},
and $\Delta\tilde{r}^{(2)}$ stands for a sum of explicit two-loop corrections
involving the top-bottom isodoublet and the Higgs boson
 (Cf. Eq.(14) of Ref.\cite{dgs}). In turn,
 $\dr^{(1)}$ can be decomposed according to 
\be
\dr^{(1)}=\Delta\alpha + \dr^{(1)}_{tb} + \dr^{(1)}_{lf} +
\dr^{(1)}_{b},\label{dr1}
\ee
where $\dr^{(1)}_{tb}$ and $\dr^{(1)}_{lf}$ denote the one-loop
contributions of the
top-bottom isodoublet and the light fermions, not contained in $\Delta\alpha$,
and $\dr^{(1)}_{b}(\mh)$ is the bosonic contribution, as defined in
Ref.\cite{si80} (i.e. including vertex parts and box diagrams).
The light fermions include the leptons and the first two generations of
quarks. At the one-loop level the $\mh$ dependence resides in
$\dr^{(1)}_{b}(\mh)$. At the two-loop level, 
the $\mh$ dependence of \equ{drexp} is contained in 
$2(\dr^{(1)} - \dr^{(1)}_{b} )\,\dr^{(1)}_{b}(\mh) +
 \Delta\tilde{r}^{(2)}(\mh)+(\dr^{(1)}_{b}(\mh))^2$.
The last term, however, does not belong to the ${\cal C}^{(2)}
(f,H)$  class and, for this reason, it is not included in the
analysis of Ref.\cite{bw}. Furthermore, it is not affected by the
HTE. Therefore, in order to test the HTE by comparing the calculations of
Refs.\cite{dgs} and \cite{bw}, we disregard 
$(\dr^{(1)}_{b}(\mh))^2$.
At the two-loop level, the contribution of \equ{drexp} to 
$(\dr_N)_{sub} (\mh)$
is then given by 
\be
(\dr_N)_{sub} (\mh)= (\dr^{(1)})_{sub} +
(\dr^{(1)})^2_{sub} -(\dr_b^{(1)})^2_{sub} +
 \Delta\tilde{r}^{(2)}_{sub}.\label{e5}
\ee
with the understanding that  $(\dr_b^{(1)})^2_{sub}$ is not included.
The corresponding contribution in the approach of Ref. \cite{bw}
is expressed as 
\be
(\dr_N)_{sub} (\mh)= (1+2\Delta\alpha)\,(\dr^{(1)})_{sub}  +
\dr^{(2),tb}_{sub} + \dr^{(2),lf}_{sub} ,\label{e6}
\ee
where the last two terms stand for the t-b isodoublet and light-fermion
contributions not contained in $\Delta\alpha$.
It should be noted that  $\dr^{(2),lf}_{sub}(\mh)$ in \equ{e6}
includes all the relevant reducible and irreducible two-loop diagrams, 
while the corresponding light fermion contribution in \equ{e5} arises
 only from the reducible terms in $(\dr^{(1)})^2_{sub}$.

In Table 1
we compare the results for $(\mw)_{sub}(\mh)$ obtained by using either 
 \equ{e5} or \equ{e6}, and we list the corresponding
shifts $\delta\mw$ in the case $\mh^0=$65 GeV (the reference point chosen in
Ref.\cite{bw}). We use the input parameters of Ref.\cite{dgs}, namely
$\mz=91.1863$ GeV, $\mt=175$ GeV, $(\Delta\alpha)_h^{(5)}=0.0280$.
In order to carry out the comparison between the two approaches in as close a
manner as possible, the two-loop contributions in both calculations are
evaluated at fixed $\mw=80.37$ GeV, and $\mw(\mh)$ is then found by using
iteratively \equ{num} in conjunction with either \equ{e5} or \equ{e6}.
(Our conclusions are very insensitive to the precise
value of $\mw$ employed in the evaluation of the two-loop corrections.)
The calculation based on 
\equ{e6} employs for $\mw(\mh^0=65$ GeV) the value obtained from the OSII 
scheme of Ref.\cite{dgs}, subject to the approximation of \equ{e5} and the
iterative method explained above. The quantities $(\mw)_{sub}(\mh)$ (second and
third column of Table 1) are obtained by subtracting $\mw(65 $ GeV). 
It is worth noting that $(\mw)_{sub}(\mh)$ is also
very insensitive to the precise value of $\mw(65 $ GeV).
The shift $\delta \mw$ represents the variation of $(\mw)_{sub}
(\mh)$ when one employs \equ{e6} relative to the value obtained from \equ{e5}.

From Table 1 we see that, over the large range 65 GeV\,$\le \mh\le1$ TeV, the
$\delta\mw$ values are very 
small, $|\delta\mw|\le 0.8$ MeV. There are a number of
significant differences between the comparison in Table 1 and those carried out
in Refs.\cite{bw,georg}: i) $(\mw)_{sub}(\mh)$, obtained in 
Refs.\cite{bw,georg} from Eqs.(\ref{num},\ref{e6}), is compared 
 with the results derived from Eqs.(\ref{num},\ref{e5}), 
rather than \equ{den}. 
In fact, the latter is a resummed expression that
includes 
terms of third and higher order involving $\Delta\alpha$. The comparison of 
Eqs.(\ref{num},\ref{e6}) and Eqs.(\ref{num},\ref{e5})
is much closer, as both expansions are truncated in second order and possible
deviations arising from different 
treatments of higher-order corrections are avoided;
 ii) As explained before, the contribution $(\dr^{(1)}_b)^2_{sub}$ is
excluded in \equ{e5}, in correspondence with \equ{e6}, as it does not belong to
${\cal C}^{(2)}(f,H)$ and is not relevant to the test of the
HTE; iii) The light fermion contribution in \equ{e6} is retained, rather than
subtracted (the consequence of excluding these contributions in both
calculations are discussed later on and in Table 3); iv) As mentioned above, in
analogy with the  treatment of \equ{e6}, the two-loop corrections in \equ{e5}
are evaluated at fixed $\mw$.

In order to extend these considerations to $\sineff$, we recall that, in the
on-shell renormalization scheme,
\be
\sineff(\mh)= k(\mh)\ s^2 \, ,
\label{e7}\ee
where $ s^2=1-\mw^2/\mz^2$,
$k(\mh)=1+\Delta\kappa$ is an \ew form factor, and $\Delta\kappa$ is an
important radiative correction.
In Eq.(17) of Ref.\cite{dgs}, $\Delta\kappa$  is parametrized in the form
\be
\Delta\kappa= \frac{8\mw^2 G_\mu}{\sqrt{2}}\left[ 
\Delta \bar{k}(s^2)+\frac{c^2}{s^2} \Delta\bar{\rho}(s^2) + \Delta\tilde{
 k}^{(2)}\right],
\label{e8}\ee
where the first two terms contain one and two-loop effects, while  the third
is an explicit reducible two-loop contribution. On the other hand, the
calculation of Ref.\cite{georg} is parametrized in terms of $\alpha$ and $
s^2$. In the on-shell scheme, physical amplitudes are frequently parametrized
in terms of $G_\mu$ and $\mw$ (or $\mz$), as this procedure prevents the
occurence of large vacuum-polarization contributions involving mass
singularities \cite{dgs,ms}. However, for the purpose of the present
comparison, which involves only subtracted quantities at the two-loop level, it
is sufficient to insert in
\equ{e8} $8\mw^2 G_\mu/\sqrt{2}=4\pi \alpha/[s^2 (1-\dr)]$,
expand $(1-\dr)^{-1}\approx 1+\dr$, and retain the additional two-loop
contribution $(4\pi\alpha/s^2) \, \dr^{(1)}[\Delta \bar{k}^{(1)} +
c^2/s^2  \Delta\bar{\rho}^{(1)}]$.
In this way, the expression based on Ref.\cite{dgs} is put in a form analogous
to that of Ref.\cite{georg}, as the latter is strictly a two-loop calculation
and employs the $(\alpha,s^2)$ parametrization.
We also subtract $(4\pi\alpha/s^2) \, \dr_{b}^{(1)}[\Delta \bar{k}^{(1)} +
c^2/s^2  \Delta\bar{\rho}^{(1)}]_{b}$, since this contribution does not 
belong to the ${\cal C}^{(2)}(f,H)$  class and is not contained in the work
of Ref. \cite{georg}.
Numerically, this term turns out to be very small. As before, in analogy with
Refs.\cite{bw,georg}, we evaluate the two-loop contributions at fixed $\mw$.
Writing \equ{e7} in the form $\sineff=s^2 +
\Delta\kappa \ s^2$, in the approach of Ref.\cite{georg} the
second term is studied considering the subtracted quantity 
$ (\Delta\kappa\ s^2 )_{sub}=
(\Delta\kappa \ s^2 )(\mh)-(\Delta\kappa \ s^2)(\mh^0)
$, and assuming that $(\Delta\kappa \ s^2)
(\mh^0)$ coincides with the value derived from the OSII scheme
of Ref.\cite{dgs}. The values of $\mw(\mh)$ in each calculation are the ones
obtained in the analysis leading to Table 1. In particular, 
$s^2=1-\mw^2/\mz^2$ in \equ{e7} is evaluated in this manner.
In Table 2 we compare the subtracted quantity 
\be
(\sineff)_{sub} (\mh)=
\sineff(\mh)-\sineff(\mh^0),
\label{e9}\ee
as evaluated in the two approches
for $\mh^0=65$ GeV. From Table 2 we see that, over the large range 65
GeV$\le\mh\le$1 TeV, the $\delta\seff$ values are very small,
$|\delta\seff|\le 1.2 \times 10^{-5}$. The results are found to be very
insensitive to the precise value of $\sineff(\mh^0)$.

In order to test the accuracy of the approximate treatment of 
${\cal C}^{(2)}(f,H)$ in Refs.\cite{dgv,dgs,dg,dgps}, which is one of our main
objectives, it is important to include in both calculations the two-loop 
effects involving the light fermions. There is also a theoretical argument that
leads to the same conclusion. In a consistent calculation 
 of $\dr_N$ at the two-loop level, one should include 
 the reducible contribution 
$2 \,\dr^{(1)}_{lf} \dr^{(1)}_{b}$ induced when one inserts \equ{dr1} into 
\equ{drexp}. In fact, without its inclusion, 
 the large contribution $2\,\Delta \alpha\, 
\dr^{(1)}_{b}$, generated by the same substitution,
 becomes somewhat arbitrary since, with equal
justification, one could separate out $\Delta \alpha+\epsilon$.
For instance, $-\epsilon$ could be the non-logarithmic part of $\Delta \alpha$,
or $\Delta \alpha+\epsilon$ could be 
the $\ms$ version of $\Delta \alpha$, or $\Delta \alpha+\epsilon$ 
could be the vacuum polarization function evaluated at $q^2\neq \mz^2$.
 With the inclusion of all the
fermionic contributions, the ambiguity disappears
since what is added to $\Delta \alpha$ must be subtracted from the remaining
fermionic component. An analogous role is played by 
$\dr^{(2)lf}_{sub}(\mh)$ in \equ{e6}, a contribution that includes both 
reducible and irreducible components. A similar observation applies 
to the calculation of $\sineff$. As explained before, when $\Delta\kappa$ 
is parametrized in terms of $\alpha$, reducible contributions proportional to 
$\dr^{(1)}$ are generated, and one must include the light fermion 
contributions in order to obtain an unambiguous answer. However, in comparing
 the two calculations it is also interesting to inquire about the specific 
difference arising from the two treatments of the $t-b$ isodoublet.
Indeed, it is in these contributions that the enhancement factors
$(\mt^2/\mw^2)^n$ ($n=1,2$) emerge at the two-loop level. A simple way of 
evaluating  this difference is to neglect the two-loop contributions involving
light fermions in both approaches and repeat the comparative analysis 
discussed before.
In Table 3 we list the corresponding $\delta\mw$ and $\delta\seff$
shifts. 
We find now $|\delta\mw|\lequiv \,1.9$ MeV and $|\delta\seff|\le 
4.5\times 10^{-5}$, with the maximal values attained at large $\mh$.
Although these shifts, arising from differences in the treatment of the
top-bottom isodoublet, are larger than the very small variations in the
complete calculations, displayed in Tables 1 and 2, they remain
small. Nonetheless, comparing Tables 1 and 2 with Table 3, we see that, at the
$O(\rm MeV)$ level in $\delta\mw$ and $O(10^{-5})$ in $\delta\seff$, the
differences in the treatment of the light fermions are significant. In fact,
their inclusion reduces the magnitude of the shifts in the complete
calculations. We also stress that, in the on-shell scheme employed in this
paper,  $\delta\mw$ and $\delta\seff$ are highly correlated. For instance, in
Table 3 at $\mh=1$ TeV, we have $\delta\mw=1.9$ MeV and this induces a change 
$\delta\seff=-3.7\times 10^{-5}$ in the tree level $s^2=1-\mw^2/\mz^2$.
Thus, the shift $\delta\seff=-4.5 \times 10^{-5}$ at $\mh=1$ TeV, shown in
Table 3, is mainly due to the effect of $\delta\mw$ in the tree level
correction to $\seff$, with only a very small change $-0.8\times 10^{-5}$
attributable to differences between the precise and HTE evaluation of the
radiative correction $s^2 \Delta \kappa$.

As a final check, in order to discriminate the effect of the iteration,
we have compared the calculations of the subtracted
radiative corrections $(\dr_N)_{sub}$ and $\Delta\kappa_{sub}$, obtained on 
the basis of Refs.\cite{bw,georg} and Ref.\cite{dgs}, when both the one and 
two-loop contributions are evaluated at fixed $\mw=80.37$ GeV (see also 
Ref.\cite{paolo}). For the 
differences between the two calculations in the range
65 GeV$\le\mh\le$1 TeV, we find
$|(\dr_N)_{sub}|\le 0.5\times 10^{-4}$ and $|\Delta\kappa_{sub}|\le
0.8 \times 10^{-4}$, which induce shifts $|\delta\mw|\le 0.8$MeV 
and $|\delta\seff|\le 1.8\times 10^{-5}$, respectively.
When the light fermions are excluded, we obtain
$|(\dr_N)_{sub}|\le 1.0\times 10^{-4}$ and $|\Delta\kappa_{sub}|\le
0.5 \times 10^{-4}$, which induce shifts $|\delta\mw|\le 1.6$MeV 
and $|\delta\seff|\le 1.1\times 10^{-5}$, respectively.
These effects are of the same order of magnitude as shown in Tables 1-3. 
However, the exact details differ, since the fixed $\mw$ calculations 
do not take into account the iterative evaluation of $\mw(\mh)$ from 
\equ{num} and, for this reason, have less physical meaning.

From Tables 1 and 2 we see that, over the large range 65 GeV$\le\mh\le$ 1 TeV,
$|\delta\mw|\le 0.8$MeV, $|\delta\seff|\le 1.2\times 10^{-5}$.
The maximal variations are larger, but still small, when the light fermion
contributions are excluded, namely $|\delta\mw|\le 1.9$MeV, 
$|\delta\seff|\le 4.5\times 10^{-5}$. We recall that the current
estimate of $\mh$  and its 95\% C.L. upper bounds are in 
the $\mh\lequiv \,300$ GeV range. From Tables 2 and 3,
 we see that in that domain
the maximum difference in $\sineff$ amounts to $-1.2\times 10^{-5}$
($-1.8\times 10^{-5}$), when the light fermion contributions are included
(excluded). 
Variations of this magnitude would induce a change of 2.3\% (3.5\%), or about
6 GeV (9 GeV), in the current  95\% C.L. upper bound $\mh\le262$ GeV.
On the other hand, the latter upper bound \cite{ewwg} 
already includes an estimated uncertainty due to higher order \ew effects 
which is significantly larger than the shifts we have just considered.
We also note that the choice $\mh^0=65$ GeV for the reference point
 is somewhat arbitrary.
Although a change in $\mh^0$ does not affect the variation of the subtracted
 quantities between one $\mh$ value and another, it does modify $|\delta\mw|$ 
and $|\delta\seff|$ at fixed $\mh$. For instance, if $\mh^0=300$ GeV were 
chosen,  $\delta\seff$ would vary from $0.8\times 10^{-5}$ to
$0.5\times 10^{-5}$ in Table 2, and  from 
$1.1\times 10^{-5}$ to
$-2.7\times 10^{-5}$ in Table 3, leading to smaller maximal values for\
$|\delta\seff|$.

We conclude our comments with an observation concerning $(\dr^{(1)}_b)^2$.
Although this contribution has not been included in our analysis, it is natural
to consider its effect in discussions of scheme dependence.
For $\mw=80.37$ GeV, $(\dr^{(1)}_b)^2$ equals $(0.04,0.11,0.53,1.05,1.57)
\times 10^{-4}$ at $\mh=(65,100,300,600,1000)$ GeV. At the $10^{-4}$ 
level of accuracy, it becomes relevant only at the higher $\mh$ values.
There exists, however, another well-known two-loop contribution involving 
$\mh$, namely the irreducible contributions proportional to $\mh^2$
\cite{veltman,barb}. Its effect on $\dr$ is $-0.98\times 10^{-4}
(\mh/\rm TeV)^2$ \cite{barb}.
Although it is usually neglected, as in the calculations of Ref.\cite{dgs},
our observation is that it would be natural to include it, together with 
$(\dr^{(1)}_b)^2$, if large values of $\mh$ are considered to test the scheme 
dependence. The combined contribution of $(\dr^{(1)}_b)^2$ and the two-loop 
effects proportional to $\mh^2$ equals (0.04,0.10,0.44,0.70,0.59)$\times 
10^{-4}$. This reduces the magnitude of these effects and the ambiguity
associated with the possible inclusion or exclusion of their contribution.
It is also worth noting that, if the $\mh$-dependence of the full two-loop
bosonic contributions is of the same magnitude as in 
 $(\dr^{(1)}_b)^2$, it would
be significantly smaller than that arising from the whole 
${\cal C}^{(2)}(f,H)$.

In summary, as illustrated in Tables 1-3, by comparing the results of 
Ref.\cite{dgs} with those of Refs.\cite{bw,georg} in the evaluation of 
subtracted quantities over the large range 65 GeV$\le\mh\le$ 1
 TeV, we have found only small differences attributable to 
the use of the HTE. When the complete calculations are compared (Tables 1 and
2) they are significantly smaller than  the errors  estimated in
Ref.\cite{dgs,dgps} at fixed $\mh$, while they reach about the same magnitude
when light fermion contributions are excluded (Table 3).
 We would like to stress, however, that these reassuring 
conclusions are not a substitute for the very difficult, but fundamental 
task, of achieving a complete two-loop evaluation of $\dr$ and other basic 
radiative corrections of the Standard Theory.

\vspace{.8cm}
We are grateful to G.~Degrassi for interesting discussions.
A.S. would like to thank the members of the Max-Planck-Institut f\"ur Physik
for their warm hospitality, and the Alexander von Humboldt Foundation
for its kind support. This research was supported in part by NSF grant 
No.PHY-9722083, by the Bundesministerium f\"ur Bildung und Forschung under
contract 06 TM 874, and by the DFG Project Li 519/2-2.

%%%%%%%%%%%%%%%%%%%%%%%%%%%%%%%% REFERENCES %%%%%%%%%%%%%%%%%%%%%%%%%%%%%%%%%%%

%%%%%%%%%%%%%%%%%%%%%%%%%%%%%%%%% TABLES %%%%%%%%%%%%%%%%%%%%%%%%%%%%%%%%%%%%%%
%%%%%%%%%%%%%%%%%%% table 1
\renewcommand{\arraystretch}{1.1}
\begin{table}
\begin{center}
\begin{tabular}{|c||c|c|c|} 
\hline 
 $\mh$ & $(\mw)_{sub}$ (Eq.5)  & $(\mw)_{sub}$ (Eq.6) & 
        $\delta\mw$  \\
(\gev)&  (MeV)& (MeV)  &  ( MeV)\\
 \hline \hline 
100 &   -23.3 & -22.9 & 0.4  \\ \hline
300 &   -96.8 & -96.0 & 0.8        \\ \hline
600 &   -149.8 & -149.7 & 0.1 \\ \hline
1000  & -188.1 & -188.8 & -0.7  \\ \hline
\end{tabular} 
\caption{\sf Comparison of $(\mw)_{sub}(\mh)$, as obtained from
 \equ{e5}, based on  Ref. \cite{dgs} (see text), and \equ{e6},
based on Ref.\cite{bw}.
The latter calculation employs  an accurate evaluation
of the contributing two-loop diagrams, i.e. it
does not apply the HTE in the two-loop
corrections. The shift $\delta\mw$ is the difference between columns 3 and 2.
}
\end{center} 
\end{table}
%\vfill 
%\newpage

%%%%%%%%%%%%%%%%%%% table 2
\renewcommand{\arraystretch}{1.1}
\begin{table}
\begin{center}
\begin{tabular}{|c||c|c|c|} 
\hline 
 $\mh$ &
 $ (\seff)_{sub}\times 10^3$&  $ (\seff)_{sub}\times 10^3$&  $\delta\seff
$\\
(\gev)&  Ref.\cite{dgs}& Ref.\cite{bw,georg}  &   $10^{-5}$  \\
 \hline \hline 
100 &   0.211 & 0.207 & -0.4  \\ \hline
300 &   0.781 & 0.769 & -1.2      \\ \hline
600 &   1.152& 1.141 & -1.1 \\ \hline
1000  & 1.412 & 1.405 & -0.7  \\ \hline
\end{tabular} 
\caption{\sf Comparison of $(\seff)_{sub}(\mh)$, as
derived from Ref.\cite{dgs} (see text), with
the corresponding results from Refs.\cite{bw,georg}.
The latter calculation is based on an accurate evaluation
of the contributing two-loop diagrams, i.e. 
does not employ the HTE in the two-loop
corrections.
The shift $\delta\seff$ is the difference of columns 3 and 2.
}
\end{center} 
\end{table} 
%%%%%%%%%%%%%%%%%%% table 3
\renewcommand{\arraystretch}{1.1}
\begin{table}
\begin{center}
\begin{tabular}{|c||c|c|c|c|} 
\hline 
 $\mh$  & 
        $\delta\mw$ & $\delta\seff$ & $(\mw)_{sub}$ &
 $ (\seff)_{sub}$ \\
(\gev)  & (MeV)& $(10^{-5})$ & (MeV) & $(10^{-3})$\\
 \hline \hline 
100 &    0.6 & -0.7 & -22.5 & 0.202 \\ \hline
300 &    1.6 & -1.8 & -94.2 & 0.748      \\ \hline
600 &    1.9 & -3.9 & -146.5 & 1.106 \\ \hline
1000  &  1.9 & -4.5 & -184.4 & 1.358 \\ \hline
\end{tabular} 
\caption{\sf The differences $\delta\mw$ and $\delta\seff$
between the calculations based on Refs.\cite{bw,georg} and Ref.\cite{dgs} 
(see text), when the light-fermion contributions are excluded in both analyses.
The results reflect the effect of applying the HTE to the two-loop
corrections involving the top-bottom isodoublet. In columns 4 and 5 we also
report the results for $(\mw)_{sub}$ and $ (\seff)_{sub}$ from the calculations
of Refs.\cite{bw,georg}. The analogous values in the approach of Ref.\cite{dgs}
can be obtained combining columns 2-4 and 3-5. 
}
\end{center} 
\end{table} 
\end{document}